\documentclass{article}
\usepackage{spconf,amsmath,graphicx,hyperref}
\usepackage{booktabs}
\usepackage{multirow}
\usepackage[table,xcdraw]{xcolor}
\usepackage{enumitem}

\title{Technical Report of Nomi Team in the \\Environmental Sound Deepfake Detection Challenge 2026}
\name{Candy Olivia Mawalim*, Haotian Zhang*, Shogo Okada \thanks{These authors contributed equally.}}
\address{Japan Advanced Institute of Science and Technology (JAIST)}
\begin{document}
%
\maketitle
\begin{abstract}
This paper presents our work for the ICASSP 2026 Environmental Sound Deepfake Detection (ESDD) Challenge. The challenge is based on the large-scale EnvSDD dataset that consists of various synthetic environmental sounds. We focus on addressing the complexities of unseen generators and low-resource black-box scenarios by proposing an audio-text cross-attention model. Experiments with individual and combined text-audio models demonstrate competitive EER improvements over the challenge baseline (BEATs+AASIST model).
\end{abstract}
\section{ESDD Challenge}
\subsection{Tracks}
The ICASSP 2026 ESDD Challenge features two tracks \cite{yin2025esdd2026environmentalsound}.
\textbf{Track 1: ESDD in Unseen Generators}:
This track aims to explore the generalizability of ESDD models to text-to-audio (TTA) and audio-to-audio (ATA) generators not seen during the training phase. The model must be robust enough to detect deepfakes created by novel or different generative systems.
\textbf{Track 2: Black-Box Low-Resource ESDD}:
This track simulates a more challenging real-world scenario characterized by extreme uncertainty and a very limited amount of data from the black-box generators (1\% of the total data). The model must effectively detect deepfakes despite minimal exposure to the generation process.

\subsection{Dataset}
The EnvSDD \cite{Yin2025_EnvSDD} used in this challenge. It consists of 45.25 hours of real sound and 316.7 hours of fake sound. The real sounds are compiled from several established public datasets, including UrbanSound8K, TAU UAS 2019 Open Dev, TUT SED 2016, TUT SED 2017, DCASE 2023 Task7 Dev, and Clotho. The fake sounds within the EnvSDD are generated using various TTA and ATA models.

\section{Proposed Method}
To achieve robust performance across both challenge tracks, we propose an architecture utilizing cross-attention with an audio-text modality. Text input is generated by an audio captioning model \cite{kadlčík2023whispertransformeraudiocaptioning}. Unlike standard speech-to-text, this transformer encoder-decoder model performs automatic audio captioning, describing the content of audio clips, such as prominent sounds or environmental noises. We extracted captions from three distinct styles: AudioSet, AudioCaps, and Clotho. Following extraction, the text representations were obtained using a text encoder before being input into our cross-attention module.

The audio input is processed at a $44.1 \text{kHz}$ sampling rate, utilizing both the raw waveform and spectrogram as initial acoustic features. The core audio processing backbone is an adaptation of the AASIST model \cite{jung2022_AASIST}. To obtain rich and generalized acoustic features, we integrate the self-supervised learning model of BEATs \cite{chen23ag_beats}.

\subsection{Text-Guided Cross-Attention Model}
To bridge the semantic gap between the acoustic and textual modalities, we employ a text-guided cross-attention model. Specifically, the acoustic features serve as Queries to retrieve relevant semantic information from the text embeddings (Keys and Values). This mechanism acts as a feature filter, enhancing audio representations that align with the textual prompts while suppressing irrelevant noise. The fused features are subsequently modeled by a stacked Gated Recurrent Unit network \cite{cho2014learning} to refine the temporal dynamics for the final detection. To train this model, we use a weighted cross-entropy loss function to address potential class imbalances between genuine and spoofed samples. The computational complexity of the model is characterized by a total of $158.91 \text{M}$ parameters during inference. Training was conducted using a single NVIDIA RTX 5080 Laptop GPU. 

\subsection{Ensemble Model}

To maximize performance and leverage multimodal information, we propose a system centered on a stacked regression ensemble combining three variants of the text-guided cross-attention model and BEATs-AASIST baseline model with features extracted from multiple pre-trained acoustic and text models. We used the text feature representation from ROBERTa (base) \cite{liu2019roberta} for the input to final ensemble meta-learner. This approach allows us to obtain benefits of the strengths of diverse model architectures while mitigating individual model weaknesses.

The final prediction is generated by an ensemble of three base systems via a stacking approach.
\begin{itemize}[leftmargin=*,nosep]
    \item Base Models (Stage 1): The primary models (including the text-guided cross-attention model variant) are trained independently to produce initial predictions.
    \item Meta-Learner (Stage 2): The predictions from the base models, combined with the RoBERTa text features are used to train a stack regression model. The meta-learner is a combination of three classic regressors: gradient boosting, random forest, and linear regressors. This stacked architecture allows the meta-learner to learn the optimal way to combine the strengths and correct the biases of the individual models, leading to a more accurate prediction.
\end{itemize}

The entire ensemble model is trained using the Mean Squared Error (MSE) as the loss function, augmented with L2 regularization to prevent overfitting. The computational complexity of the system is substantial, with a total of $284.64 \text{M}$ parameters involved during inference. Training was executed on a single high-performance Quadro RTX A6000 GPU.

\section{Experiments and Results}

\subsection{Progress phase}

Initial experiment in progress phase indicates that our proposed ATCA (Audio-Text Cross-Attention) model shows promising results on the Track 2 Evaluation set, achieving an EER of \textbf{12.03\%} with only 10\% data used for training and validation. This result is lower than the BEATS+AASIST baseline EER of 12.64\% for the evaluation set.

\subsection{Ranking phase}
Table~\ref{tab:track1_results} presents the ranking phase results for Track 1 (ESDD in Unseen Generators), while Table~\ref{tab:track2_results} shows the corresponding results for Track 2 (Black-Box Low-Resource ESDD). Our individual ATCA model demonstrates strong performance, yielding a lower EER compared to the baseline BEATs+AASIST model in Track 1. However, the EER achieved by ATCA is similar to the baseline in Track 2. This observation suggests that the benefit of the text modality may be diminished in the Track 2 test set, possibly due to poorer text recognition or lower quality captions generated for the specific, low-resource audio data.

The utilization of the ensemble model reduces the EER. This improvement is achieved by combining the strengths of the individual models. Unfortunately, this performance gain comes at the cost of increased computational complexity, as the total number of parameters increases proportionally to the individual models incorporated in the ensemble.

\begin{table}[h]
    \centering
    \caption{Ranking phase results in EER (\%) for Track 1}
    \resizebox{0.8\columnwidth}{!}{
    \begin{tabular}{|l|c|cc|}
    \hline
    \rowcolor[HTML]{F3F3F3} 
    \cellcolor[HTML]{F3F3F3} & \cellcolor[HTML]{F3F3F3} & \multicolumn{2}{c|}{\cellcolor[HTML]{F3F3F3}\textbf{EER (\%)}} \\ \cline{3-4} 
    \rowcolor[HTML]{F3F3F3} 
    \multirow{-2}{*}{\cellcolor[HTML]{F3F3F3}\textbf{System}} & \multirow{-2}{*}{\cellcolor[HTML]{F3F3F3}\textbf{Param (M)}} & \multicolumn{1}{c|}{\cellcolor[HTML]{F3F3F3}\textbf{Validation}} & \textbf{Test} \\ \hline \hline
    AASIST & 0.30 & \multicolumn{1}{c|}{0.92} & 15.02 \\ \hline
    BEATs+AASIST & 90.73 & \multicolumn{1}{c|}{0.10} & 13.20 \\ \hline
    ATCA & 158.91 & \multicolumn{1}{c|}{0.10} & 11.28 \\ \hline
    ATCA-ens & 284.64 & \multicolumn{1}{c|}{0.07} & 11.22 \\ \hline
    \end{tabular}}
    \label{tab:track1_results}
    \vspace{-6mm}
\end{table}

\begin{table}[h]
    \centering
    \caption{Ranking phase results in EER (\%) for Track 2}
    \resizebox{0.8\columnwidth}{!}{
    \begin{tabular}{|l|c|cc|}
    \hline
    \rowcolor[HTML]{F3F3F3} 
    \cellcolor[HTML]{F3F3F3} & \cellcolor[HTML]{F3F3F3} & \multicolumn{2}{c|}{\cellcolor[HTML]{F3F3F3}\textbf{EER (\%)}} \\ \cline{3-4} 
    \rowcolor[HTML]{F3F3F3} 
    \multirow{-2}{*}{\cellcolor[HTML]{F3F3F3}\textbf{System}} & \multirow{-2}{*}{\cellcolor[HTML]{F3F3F3}\textbf{Param (M)}} & \multicolumn{1}{c|}{\cellcolor[HTML]{F3F3F3}\textbf{Validation}} & \textbf{Test} \\ \hline \hline
    AASIST & 0.30 & \multicolumn{1}{c|}{1.11} & 15.40 \\ \hline
    BEATs+AASIST & 90.73 & \multicolumn{1}{c|}{0.16} & 12.48 \\ \hline
    ATCA & 158.91 & \multicolumn{1}{c|}{0.15} & 12.64 \\ \hline
    ATCA-ens & 284.64 & \multicolumn{1}{c|}{0.07} & 11.98 \\ \hline
    \end{tabular}}
    \label{tab:track2_results}
    \vspace{-6mm}
\end{table}

\section{Conclusion}
We proposed an audio-text cross-attention model for the ESDD Challenge 2026. The initial development and parameter optimization of the proposed model demonstrate the importance of the integration of semantic text, especially against unseen generators (Track 1). Furthermore, the stacked regression ensemble model gives further improvement by addressing the challenges of low-resource data (Track 2).


\bibliographystyle{IEEEbib}
\bibliography{mybib}

\end{document}